\documentclass[]{aa} 

\usepackage{graphicx}
\usepackage{lscape}
\usepackage{txfonts}
\newcommand{\bfxi}{\mbox{\boldmath$\xi$}}
\newcommand{\bftau}{\mbox{\boldmath$\tau$}}
\newcommand{\bfkappa}{\mbox{\boldmath$\kappa$}}
\newcommand{\bfv}{\mbox{\boldmath$v$}}

\newcommand{\bfbigtriangledown}{\mbox{\boldmath$\bigtriangledown$}}

\begin{document}

\title{Incompressible MHD modes in the thin magnetically twisted flux tube}


   \author{O.K. Cheremnykh
          \inst{1}, 
          V. Fedun\inst{2},
          A. N. Kryshtal\inst{1},
          \and
          G. Verth\inst{3}
          }

   \institute{Department of Space Plasma, Space Research Institute, Kiev, Ukraine\\
              \email{oleg.cheremnykh@gmail.com}, 
              \and
             Department of Automatic Control and Systems Engineering, University of Sheffield, Sheffield, S1 3JD, UK\\
             \email{v.fedun@sheffield.ac.uk}, 
             \and
            School of Mathematics and Statistics, University of Sheffield, Sheffield, S3 7RH, UK\\
             \email{g.verth@sheffield.ac.uk},
             }

\titlerunning{MHD modes in the thin magnetically twisted flux tube}
\authorrunning{O.K. Cheremnykh et al.}

   \date{Received XXXX}
   
 
  \abstract
   {Observations have shown that twisted magnetic fields naturally occur, and indeed  are omnipresent in the Sun's atmosphere. It is therefore of great theoretical interest in solar atmospheric waves research to investigate the types of magnetohydrodynamic (MHD) wave modes that can propagate along twisted magnetic flux tubes.}
   {Within the framework of ideal MHD, the main aim of this work is to investigate small amplitude incompressible wave modes of twisted magnetic flux tubes with $m \ge 1$. The axial magnetic field strength inside and outside the tube will be allowed to vary, to ensure the results will not be restricted to only cold plasma equilibria conditions.}
   {The dispersion equation for these incompressible linear MHD wave modes was derived analytically by implementing the long wavelength approximation.}
   {It is shown, in the long wavelength limit, that both the frequency and radial velocity profile of the $m=1$ kink mode are completely unaffected by the choice of internal background magnetic twist. However, fluting modes with $m \ge 2$ are sensitive to the particular radial profile of magnetic twist chosen. Furthermore, due to background twist, a low frequency cut-off is introduced for fluting modes that is not present for kink modes. From an observational point of view, although magnetic twist does not affect the propagation of long wavelength kink modes, for fluting modes it will either work for or against the propagation, depending on the direction of wave travel relative to the sign of the background twist.}
   {}


\maketitle
%

\section{Introduction}
An axially symmetric, vertical and magnetically twisted flux tube is a convenient model for analytical studies of various magnetohydrodynamic (MHD) perturbations. For a long time this approximation was the focus of investigations of MHD wave propagation in solar and space plasmas  \citep[see e.g.][]{Priest1991, Bennett1999, Erdelyi2007, Ruderman2007, Ladikov-Roev2013, Cheremnykh2014} 
and high-temperature \citep[see e.g.][]{Suydam1959, Bateman1978, Sudan1989, 
Cheremnykh1989, Burdo1994} plasmas. This geometry is also a useful approximation in solving
fundamental problems of plasma physics \citep[see for example][]{Trubnikov1966, Filippov2007, 
Cheremnykh1994, Zagorodny2014}, to name but a few. 
In spite of many previous theoretical studies of wave propagation in solar magnetic flux tubes many questions still remain open. There are at least two contradictory opinions on how the radial dependence of the equilibrium azimuthal component of the magnetic field outside of the flux tube should be modelled. \citet[][]{Filippov2007, Vrsnak2008} assume that external magnetic field decreases with distance from the tube boundary inversely proportional to the radius, that is, as a function of $1/r$.  This approximation was previously used by, for example,  \citet[][]{Erdelyi2006, Ruderman2015, Giagkiozis2015}. Recently, \citet{Giagkiozis2015} has shown that the wave solution for a background constant twist outside the tube is actually very close to the solution when the twist is proportional to $1/r$. From another point of view, by taking into account plasma conductivity, the external magnetic field does not penetrate significantly through the tube boundary and, therefore, can be neglected \citep[see e.g.][]{Parker2007, Solovev2011, Solovev2012}. In these papers authors applied simple electromagnetic assumptions confirming the absence of azimuthal components of the magnetic field outside the tube.

In the present work, we will examine MHD wave propagation in a magnetic flux tube with an internal twist only. To go beyond cold plasma equilibria conditions the axially aligned magnetic field inside and outside the flux tube are allowed to be different. A similar background model has been used previously by \citet[][]{Bennett1999} and thereafter in a number of other papers, for example \citet[][]{Erdelyi2007, Erdelyi2010}. 
In the framework of ideal MHD, we will assume incompressible linear perturbations and implement the thin tube approximation. Also, we will focus on the analytical solutions related to modes with only $m\ge1$.  The analytical dispersion relation and expression for eigenfunctions will be obtained by assuming the small parameter $\varepsilon=k_z a \ll 1$, where $a$ is the radius of the magnetic flux tube and $k_z$ is the longitudinal wavenumber. 

\section{Derivation of linear radial component wave equation}

We proceed from the linearised ideal MHD equations for the displacement vector $\bfxi$ of a finite volume element. By assuming the time dependence of all perturbed physical quantities as $exp \left( -i\omega t \right)$, these equations can be written as \citep[see e.g.][]{Priest1982, Kadomtsev1966, Zagorodny2014} 
\begin{equation}
\rho\frac{\partial^2\bfxi}{\partial t^2}=-\rho\omega^2\bfxi=\mathbf{F}\left(\bfxi \right), \label{a1}
\end{equation}%
where
   \begin{eqnarray}
 \mathbf{F}\left(\mathbf{\bfxi}\right)=-\bfbigtriangledown\delta p_1+\left(\mathbf{B}\cdot\mathbf{\nabla}\right)\delta\mathbf{B}+
\left(\delta\mathbf{B}\cdot \mathbf{\nabla}\right)\mathbf{B}, \nonumber \\
\delta p_1 = \delta p +\mathbf{B}\cdot\delta \mathbf{B} = -\gamma p \mathbf{\nabla}\cdot \bfxi-B^2\left(\mathbf{\nabla}\cdot \bfxi_{\perp} +2 \bfkappa\cdot \bfxi_{\perp}\right), \nonumber \\
\delta p =-\bfxi \cdot \bfbigtriangledown p - \gamma p \mathbf{\nabla}\cdot \bfxi, \label{a2}  \\
\delta\mathbf{B}=\bfbigtriangledown \times \left[\bfxi \times \mathbf{B}\right], \nonumber \\
\mathbf{\nabla}\cdot\delta\mathbf{B}=0, \nonumber \\
\bfkappa=\left(\bftau \cdot \bfbigtriangledown\right)\bftau, \nonumber \\
\bftau=\mathbf{B}/B. \nonumber
\end{eqnarray}
Here, the symbol $\delta$ corresponds to the perturbed quantities, $\rho$ is the equilibrium plasma density,  $p$ is the  equilibrium plasma pressure, $\bfxi=\xi_r\mathbf{e}_r+\xi_{\varphi}\mathbf{e}_{\varphi}+\xi_z\mathbf{e}_z$ is the displacement vector (where $\mathbf{e}_r$, $\mathbf{e}_\varphi$ and $\mathbf{e}_z$ are unit vectors of cylindrical coordinates $r$, $\varphi$ and $z$), $\gamma$ is the adiabatic index, $\omega$ is the angular frequency, 
$\mathbf{B}$ is the equilibrium magnetic field, $\bftau$ is the normalised magnetic field, $\delta p_1$ is the perturbation of total plasma pressure (plasma plus magnetic), and $\bfkappa$ is the vector of curvature of magnetic field lines. 
The derivation of the expression for $\delta p_1$ is shown in Appendix A. $\bfxi_\perp=\bfxi-\xi_{\parallel}\bftau$ in Eq. \ref{a2} corresponds to the perpendicular component of $\bfxi$ to the equilibrium magnetic field.
All physical quantities depend on the radial coordinate $r$. For simplicity,  the magnetic field is normalised as $\mathbf{B}/\sqrt{4\pi}\rightarrow \mathbf{B}$.   
   \begin{figure}
   \centering
   \includegraphics[width=7cm]{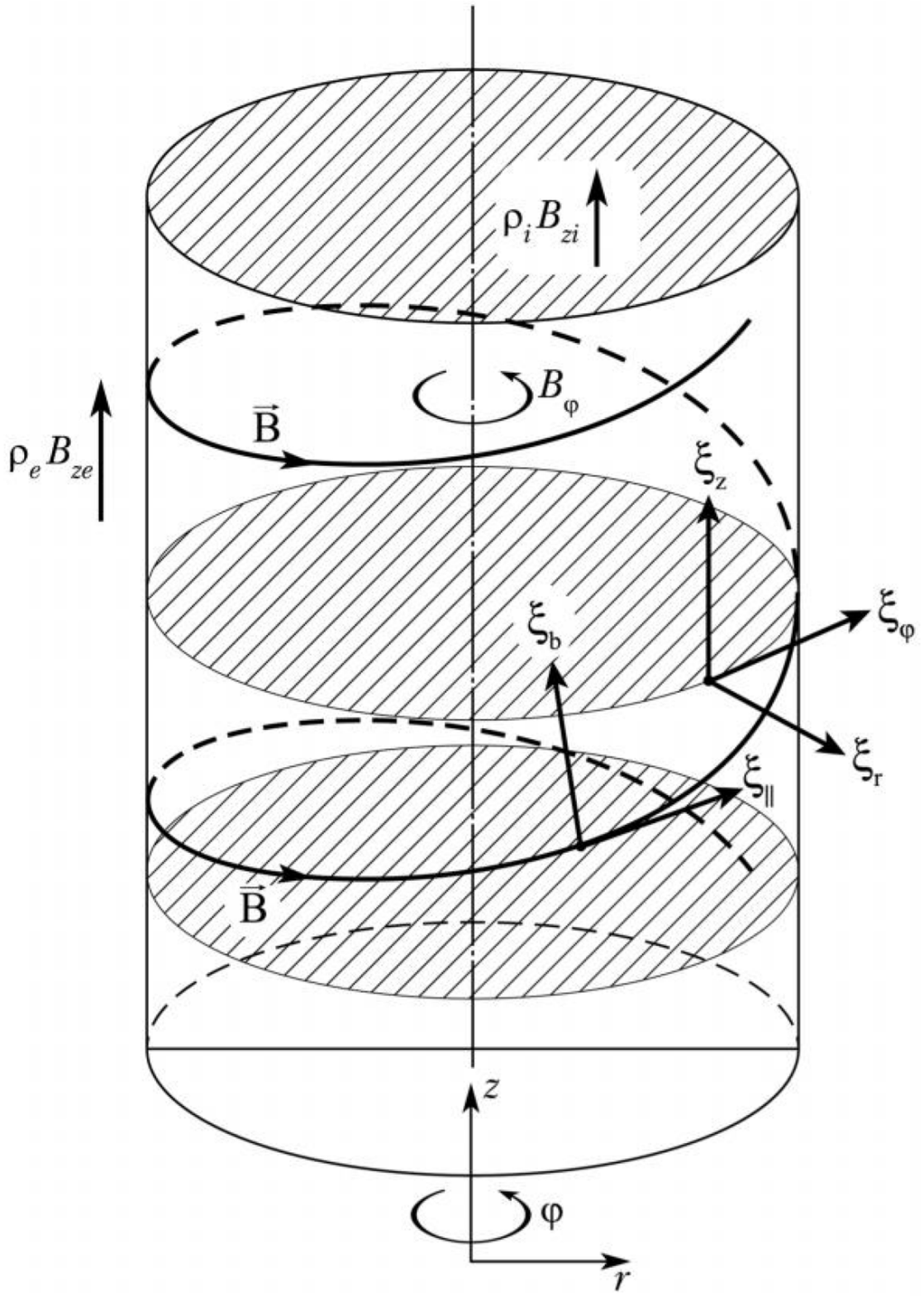}
      \caption{Magneto-hydrostatic equilibrium of the twisted magnetic flux tube.}
         \label{Fig1}
   \end{figure}
We use the cylindrical coordinate system $\left(r, \varphi, z \right)$, hence the magnetic surfaces are nested cylinders of radius $r$ and the unperturbed tube axis is parallel to the $z$-axis (see Fig. \ref{Fig1}). Both inside and outside the tube the equilibrium magnetic field is given by
\begin{equation}
\mathbf{B}=B_{\varphi}(r)\mathbf{e}_\varphi+B_z(r)\mathbf{e}_z, \label{a3}
\end{equation}
which must satisfy the magneto-hydrostatic equilibrium:
\begin{equation}
\frac{d}{dr}\left(p+\frac{B^2}{2}\right)+\frac{B_{\varphi}^2}{r}=0. \label{a4}
\end{equation}
Using the relation
\begin{equation*}
\bfkappa=\left(\frac{\mathbf{B}}{B}\cdot \bfbigtriangledown\right)\frac{\mathbf{B}}{B}=-\frac{\mathbf{e}_r}{r}\left(\frac{B_{\varphi}}{B}\right)^2,
\end{equation*}
and the rules for differentiating unit vectors, $\partial \mathbf{e}_r /\partial \varphi =\mathbf{e}_{\varphi}$ and $\partial \mathbf{e}_{\varphi}/\partial \varphi=-\mathbf{e}_r $, Eqs. (\ref{a1}) and (\ref{a2}) reduce to: 
\begin{eqnarray}
\rho\omega^2\xi_{r}-\frac{d}{dr}\delta p_1 -\frac{2B_{\varphi} \delta B_{\varphi}}{r}+i\left(\frac{m}{r}B_{\varphi}+k_{z}B_{z}\right)\delta B_r=0, \label{a5} \\
\rho\omega^2\xi_{\varphi}-\frac{im}{r}\delta p_1 +\frac{\delta B_{r}}{r}\frac{d}{dr}\left(r B_{\varphi}\right)+i\left(\frac{m}{r}B_{\varphi}+k_{z}B_{z}\right)\delta B_{\varphi}=0, \label{a6} \\
\rho\omega^2\xi_{z}-ik_{z}\delta p_{1} +\delta B_{r}\frac{d}{dr}B_{z}+i\left(\frac{m}{r}B_{\varphi}+k_{z}B_{z}\right)\delta B_{z}=0. \label{a7}
\end{eqnarray} 
To obtain Eqs. (\ref{a5}) - (\ref{a7}) we assumed that all equilibrium quantities depend on $r$ alone. Therefore, we can Fourier decompose the solution as
\begin{equation}
\bfxi\left(\mathbf{r}, t\right)=\bfxi(r)exp i \left(-\omega t+m\varphi +k_z z\right), \nonumber
\end{equation}
where $m$ is the azimuthal wave number.
In this study we shall not deal with the axi-symmetric case when $m=0$, which describes the sausage mode. Instead, we focus on all the modes with $m \ge 1$, corresponding to non-axially symmetric oscillations which are the kink $m=1$ and surface $m>1$ modes. A cartoon  of the model geometry is shown in Fig. (\ref{Fig1}).

Since the displacement vector $\bfxi(r)$
is a function of the radius alone, then the problem becomes one-dimensional. 
For more convenient analysis of the perturbations in the Eqs. (\ref{a5}) - (\ref{a7}) we changed  the $\varphi$ and $z$ components of the displacement vector $\bfxi$ and wave vector $\mathbf{k}$ to the components directed along the bi-normal (subscript $b$) and along the magnetic field lines (subscript $\parallel$):
\begin{eqnarray}
&& \bfxi=\xi_r \mathbf{e}_r+\xi_b \mathbf{e}_b+\xi_\parallel \mathbf{e}_\parallel, \nonumber \\
&&\xi_b=\xi_{\varphi}\frac{B_z}{B}-\xi_z\frac{B_{\varphi}}{B}, \quad  \mathbf{e}_b=\mathbf{e}_\varphi \frac{B_z}{B}-\mathbf{e}_z\frac{B_{\varphi}}{B}, \label{a10} \\
&& \xi_{\parallel}=\xi_{\varphi}\frac{B_{\varphi}}{B}+\xi_{z}\frac{B_z}{B}, \quad  \mathbf{e}_{\parallel}=\mathbf{e}_{\varphi}\frac{B_{\varphi}}{B}+\mathbf{e}_z\frac{B_z}{B}, \nonumber \\
&& k_{b}=\frac{m}{r}\frac{B_z}{B}-k_{z}\frac{B_\varphi}{B}, \nonumber \\
&& k_{\parallel}=\mathbf{k}\cdot\mathbf{e}_{\parallel}=\frac{m}{r}\frac{B_{\varphi}}{B}+k_{z}\frac{B_{z}}{B}. \nonumber 
\end{eqnarray}
In these more convenient variables, the components of the perturbed magnetic field can be obtained from Eq. (\ref{a2}):
\begin{eqnarray}
&& \delta B_{r}=ik_{\parallel}B\xi_r, \nonumber \\
&& \delta B_{\varphi}= ik_{z}\left(\xi_{\varphi}B_{z}-\xi_{z}B_{\varphi}\right)-\frac{d}{dr}\left(\xi_r B_{\varphi}\right), \label{a8} \\
&& \delta B_{z}=\frac{im}{r}\left(\xi_{z}B_{\varphi}-\xi_{\varphi}B_{z}\right)-\frac{1}{r}\frac{d}{dr}\left(r\xi_{r}B_z\right), \nonumber 
\end{eqnarray}
and Eqs. (\ref{a5})-(\ref{a7}) can be written as
\begin{eqnarray}
&& \left(\rho\omega^2-k_{\parallel}^2 B^2\right)\xi_{r}-\frac{d}{dr}\delta p_1-2ik_{\parallel}\frac{B_{\varphi}B_z}{r}\xi_{b}+ \nonumber \\
&& 2ik_{b}\frac{B_{\varphi}^2}{r}\xi_{b}+2\frac{B_{\varphi}}{r}\frac{d}{dr}\left(\xi_{r}B_{\varphi}\right)=0, \label{a11} \\
&& \left(\rho\omega^2-k_{\parallel}^2 B^2\right)\xi_{b}-ik_{b}\delta p_1+2ik_{\parallel}\frac{\xi_r}{r}B_{\varphi}B_{z}=0, \label{a12} \\
&& \left(\rho\omega^2-k_{\parallel}^2\frac{\gamma p}{1+\beta}\right)\xi_{\parallel}-ik_{\parallel}\frac{\beta}{1+\beta}\left[\delta p_1 -2\frac{B_{\varphi}^2\xi_{r}}{r}\right]=0, \label{a13}
\end{eqnarray}
where the values of the total pressure $\delta p_1$ and $\mathbf{\nabla}\cdot \bfxi$ are also represented via radial $\left(\xi_r\right)$, bi-normal $\left(\xi_b\right)$ and field aligned $\left(\xi_{\parallel}\right)$ displacements:
\begin{eqnarray}
\delta p_1 = -\left(\gamma p+B^2\right)\mathbf{\nabla}\cdot\bfxi+iB^2k_{\parallel}\xi_{\parallel}+2B_{\varphi}^2\frac{\xi_{r}}{r}, \nonumber \\
\mathbf{\nabla}\cdot\bfxi=\frac{1}{r}\frac{d}{dr}\left(r\xi_{r}\right)+ik_{b}\xi_{b}+ik_{\parallel}\xi_{\parallel}. \label{a14} 
\end{eqnarray}
Eqations (\ref{a11}) - (\ref{a14}) are the starting point for further analysis.
Let us reduce the Eqs. (\ref{a11})-(\ref{a13}) to the one single equation for the radial component of the displacement vector $(\xi_r)$. 
From Eqs. (\ref{a12})-(\ref{a13}) one can obtain
\begin{eqnarray}
&& k_{b}\xi_{b}=\frac{i}{\rho\left(\omega^2-\omega_{A}^2\right)}\left[k_{b}^2\delta p_1-2k_{\parallel}k_{b}\frac{\xi_r}{r}B_{\varphi}B_{z}\right], \label{a15} \\
&& k_{\parallel}\xi_{\parallel}=\frac{ik_{\parallel}^2}{\rho\left(\omega^2-\omega_{T}^2\right)}\frac{c_{T}^2}{c_{A}^2}\left[\delta p_1-\frac{2B_{\varphi}^2}{r}\xi_{r}\right]. \label{a16}
\end{eqnarray}
Here, 
\begin{eqnarray}
&& \omega_{A}^2=k_{\parallel}^2c_{A}^2, \quad \omega_{S}^2=k_{\parallel}^2c_{S}^2, \quad \omega_{T}^2=k_{\parallel}^2c_{T}^2, \nonumber \\
&& c_{A}^2=\frac{B^2}{\rho}, \quad c_{S}^2=\frac{\gamma p}{\rho}, \quad c_{T}^2=\frac{c_{S}^2}{1+\beta}, \quad \beta=\frac{c_{S}^2}{c_{A}^2}. \nonumber
\end{eqnarray}
By substituting Eqs. (\ref{a15}) and (\ref{a16}) into Eq. (\ref{a14}), we can obtain
\begin{eqnarray}
&& \delta p_1 =\frac{1}{k_{b}^2+\chi^2}\left[\rho\left(\omega^2-\omega_{A}^2\right)\frac{1}{r}\frac{d}{dr}\left(r\xi_r\right)+ \right. \nonumber \\
&& \left.2k_{\parallel}k_{b}\frac{\xi_r}{r}B_{\varphi}B_z+\frac{2B_{\varphi}^2}{r}\xi_r\chi^2\right], \label{a17} 
\end{eqnarray}
where
\begin{equation}
\chi^2=\frac{\left(\omega^2-\omega_{A}^2\right)\left(\omega_{S}^2-\omega^2\right)}{\left(c_{S}^2+c_{A}^2\right)\left(\omega^2-\omega_{T}^2\right)}. \nonumber \\
\end{equation}
From Eqs. (\ref{a4}), (\ref{a11}), (\ref{a15}) and (\ref{a17}) we obtain the governing wave equation for the linear radial component $(\xi_r)$:
\begin{eqnarray}
&& \frac{d}{dr}\left[\frac{\rho\left(\omega^2-\omega_{A}^2\right)}{k_{b}^2+\chi^2}\frac{1}{r}\frac{d}{dr}\left(r\xi_r\right)\right]+
2r\xi_r\frac{d}{dr}\left[\frac{B_{\varphi}^2}{r^2}\frac{\chi^2}{k_{b}^2+\chi^2}+\right. \nonumber \\
&& \left.\frac{B_{\varphi}B_z}{r^2}\frac{k_{\parallel}k_{b}}{k_{b}^2+\chi^2}\right]=\rho\left(\omega^2-\omega_{A}^2\right)\xi_r+2\xi_rB_{\varphi}\frac{d}{dr}\left(\frac{B_{\varphi}}{r}\right)- \nonumber \\
&& 4\xi_r\frac{B_{\varphi}^2}{r^2\rho}\frac{\chi^2}{k_{b}^2+\chi^2}\frac{\left(k_{\parallel}B_z-k_bB_{\varphi}\right)^2}{\left(\omega^2-\omega_{A}^2\right)}. \label{a18} 
\end{eqnarray}In \citet{Cheremnykh2015} it was shown that Eq. (\ref{a18}) is equivalent to the well
known Hain-L\"{u}st equation \citep[see e.g.][]{Hain1958} and also it was shown that from this  equation we can obtain Suydam's criterion \citep[see e.g.][]{Suydam1959} and stability criterion
for ballooning modes. 

\section{Incompressible perturbations in the long wavelength approximation} 

Let us assume that plasma perturbation is incompressible, that is the velocity perturbation $\delta \bfv =\partial \bfxi/\partial t$ is very small relative to the sound speed,  $c_{S}^2\rightarrow\infty$ $\left(\gamma\rightarrow \infty\right)$. It follows from the second Eq. of (\ref{a14}) and Eqs. (\ref{a15}) - (\ref{a16}) that
\begin{equation}
\mathbf{\nabla}\cdot{\bfxi}=\frac{1}{r}\frac{d}{dr}\left(r\xi_{r}\right)-\frac{\delta p_1}{\rho \left(\omega^2-\omega_{A}^2\right)}\left(k_{z}^2+\frac{m^2}{r^2}\right)+2k_{\parallel}\frac{m}{r^2}\frac{B_{\varphi}B\xi_r}{\rho\left(\omega^2-\omega_{A}^2\right)}. \label{a19}
\end{equation}
For an incompressible perturbation the expression (\ref{a17}) for $\delta p_1$ becomes
\begin{equation}
\delta p_1=\frac{1}{k_{z}^2+\left(m^2/r^2\right)}\left[\rho\left(\omega^2-\omega_{A}^2\right)\frac{1}{r}\frac{d}{dr}\left(r\xi_r\right)+2k_{\parallel}\frac{m}{r}\frac{\xi_r}{r}B_{\varphi}B\right]. \label{a20}
\end{equation}
By substituting Eq. (\ref{a20}) into Eq. (\ref{a19}), we obtain the condition for the divergence of an incompressible flow, that is,
\begin{equation}
\mathbf{\nabla}\cdot{\bfxi}=0. \label{a21} \\
\end{equation}

Eq. (\ref{a21}) can also be obtained in another way. By substituting $\delta p_1$ from Eq. (\ref{a14}) into Eq. (\ref{a13}),
we obtain $i\omega^2\xi_{\parallel}=k_{\parallel}c_{S}^2\mathbf{\nabla}\cdot \bfxi$.
By assuming $c_{S}^2\rightarrow \infty$, we also arrive at the same Eq. (\ref{a21}).

For incompressible perturbations Eq. (\ref{a18}) reduces to the equation
\begin{eqnarray}
&& \frac{d}{dr}\left[ \frac{\rho\left(\omega^2-\omega_{A}^2\right)}{k_{z}^2+m^2/r^2} \frac{1}{r} \frac{d}{dr} \left(r\xi_r\right)\right]+2r\xi_r
\frac{d}{dr}\left[\frac{BB_{\varphi}}{r^2}\frac{k_{\parallel}\left(m/r\right)}{k_{z}^2+m^2/r^2}\right]- \nonumber \\
&& \xi_r\left[\rho\left(\omega^2-\omega_{A}^2\right)+2B_{\varphi}\frac{d}{dr}\left(\frac{B_{\varphi}}{r}\right)- \right. \nonumber \\
&& \left. \frac{4\left(B_{\varphi}^2/r^2\right)}{k_{z}^2+m^2/r^2}\frac{k_z^2\omega_{A}^2}{\left(\omega^2-\omega_{A}^2\right)}\right]=0 \label{a22}
\end{eqnarray}
and coincides with the Eq. (14.36) in \citet{Miyamoto2005}. Equation (\ref{a22}) has only one singular point at Alfv{\'e}n frequency, when $\omega^2=\omega_{A}^2$. To obtain this equation we assumed that $k_{\parallel}c_{S} \rightarrow \infty$ and, therefore, the longitudinal component wave vector $k_{\parallel}$ cannot vanish.

For perturbations with a small azimuthal wave number $m$, Eq. (\ref{a22}) can be simplified further by implementing the long wavelength approximation, that is, for the case when $\varepsilon = k_z a \ll 1$. In this limit, for a homogeneous longitudinal magnetic field, $B_z=const$, Eq. (\ref{a22}) is given by
\begin{eqnarray}
&& \frac{1}{r}\frac{d}{dr}\left[\left(\rho\omega^2-F^2\right)r\frac{d\phi}{dr}\right]+\frac{dF^2}{dr}\frac{\phi}{r}-
\left(\rho\omega^2-F^2\right)\frac{m^2\phi}{r^2}+ \nonumber \\
&& 4\frac{B_{\varphi}^2}{r^2}\frac{k_{z}^2F^2}{\left(\rho\omega^2-F^2\right)}\phi=0, \label{a23}
\end{eqnarray}
where
\begin{equation}
\phi=r\xi_{r}, \,  F\left(r\right)=\frac{m}{r}B_{\varphi}\left(r\right)+k_{z}B_{z}. \nonumber
\end{equation}
To obtain Eq. (\ref{a23}) we also used the relevant long wavelength approximation that
\begin{equation}
m^2+k_{z}^2r^2=m^2+\left(\frac{r}{a}\right)^2\varepsilon^2\approx m^2. \nonumber
\end{equation}

Further analysis of  the Eq. (\ref{a23}) requires an equilibrium model for the flux tube and the boundary conditions for the perturbed quantities on its surface. 

\section{Boundary conditions}

To obtain the first boundary condition we used the equation of incompressibility given by Eq. (\ref{a21}) and relation Eq. (\ref{a14}). By following the methodology presented in \citet{Jackson1998, Priest1991}, let us integrate Eq. (\ref{a21}) over a small interval $2\delta$ near flux tube surface (i.e, within the layer $a-\delta$ and $a+\delta$, $a$ is the tube radius, $\delta \ll a$) results in
\begin{eqnarray}
&& \int_{a-\delta}^{a+\delta}r\mathbf{\nabla}\cdot\bfxi dr=
\int_{a-\delta}^{a+\delta}\left[\frac{d}{dr}\left(r\xi_r\right)+ir\left(k_{b}\xi_{b}+k_{\parallel}\xi_{\parallel}\right)\right]dr= \nonumber \\ 
&& r\xi_r\left|_{a-\delta}^{a+\delta}\right. +i\int_{a-\delta}^{a+\delta}r\left(k_{b}\xi_{b}+k_{\parallel}\xi_{\parallel}\right)dr=0. \nonumber
\end{eqnarray}
By taking into account the continuity of the integrated function, this equation can be represented as
\begin{equation}
\phi\left|_{a-\delta}^{a+\delta}\right.+2i\delta\left[r\left(k_{b}\xi_{b}+k_{\parallel}\xi_{\parallel}\right)\right]\left|_{r=a}\right.=0. \nonumber
\end{equation}
Assuming that $\delta\rightarrow 0 $, the first boundary condition can be represented in the form:
\begin{equation}
\left<\phi\right>=\phi\left(a+0\right)-\phi\left(a-0\right)=0. \label{a25}
\end{equation}
This condition requires that the radial plasma displacement is continuous at $r=a$.
Now let us obtain the second boundary condition. 
Equation (\ref{a23}) can be rewritten as 
\begin{eqnarray}
&& \frac{d}{dr}\left[\left(\rho\omega^2-F^2\right)r\frac{d\phi}{dr}\right]+\phi\frac{d}{dr}\left[\frac{2m}{r}B_{\varphi}F-m^2\left(\frac{B_{\varphi}}{r}\right)^2\right]- \nonumber \\
&& \frac{\phi}{r}\left[m^2\left(\rho \omega^2-F^2\right)-\frac{4B_{\varphi}^2k_{z}^2F^2}{\left(\rho\omega^2-F^2\right)}\right]. \label{a26} 
\end{eqnarray}
By taking into account that the magnetic field and  plasma equilibrium parameters are different inside and outside of the magnetic flux tube, Eq. (\ref{a26}) 
has different solutions for $r>a$ and $r<a$. The inner and outer solutions should agree for values of {$r$} in an intermediate region $\left(a-\delta, \, a+\delta\right)$ if $\delta \rightarrow 0$. By assuming that radial plasma displacement $\xi_r$ is continuous, the matching condition at the boundary  \citep[see e.g.][]{Soloviev1975, Priest1991} can be obtained by integrating Eq. (\ref{a26}) between $a-\delta$ and $a+\delta$:
\begin{equation}
\left<\left(\rho\omega^2-F^2\right)r\frac{d\phi}{dr}+\frac{2m}{r}B_{\varphi}F\phi-
m^2\left(\frac{B_{\varphi}}{r}\right)^2\phi\right>=0. \label{a27}
\end{equation}
For the function $\phi$ inside and outside the flux tube, Eq. (\ref{a27}) represents dispersion the relation for MHD oscillations.
According to Eq. (\ref{a20}) the total pressure perturbation is given by
\begin{equation}
\delta p_{1}=\frac{1}{m^2}\left[\frac{2m}{r}B_{\varphi}F\phi+\left(\rho\omega^2-F^2\right)r\frac{d\phi}{dr}\right]. \nonumber
\end{equation}
This means that Eq. (\ref{a27}) is equivalent to 
\begin{equation}
\left<\delta p_{1}-\frac{B_{\varphi}^2}{r^2}\phi\right>=0. \label{a27a}
\end{equation}
Equation (\ref{a27a}) is the dynamic boundary condition that is often applied to study MHD perturbations of magnetic flux tubes (\citet[][]{Erdelyi2006, Bennett1999, Erdelyi2007}, to name but a few). 

\section{General dispersion relation for $m \ge 1$ modes}
In this section, Eq. (\ref{a23}) together with the boundary conditions Eqs. (\ref{a25}) and (\ref{a27a}) will be used to obtain the dispersion equation governing oscillations of the magnetic tube. In the long wavelength approximation in Eq (\ref{a23}) the last term is proportional to $\varepsilon^2$ and hence a good approximation can be neglected, resulting in 
\begin{eqnarray}
\frac{1}{r}\frac{d}{dr}\left[\left(\rho\omega^2-F^2\right)r\frac{d\phi}{dr}\right]+\frac{dF^2}{dr}\frac{\phi}{r}-
\frac{m^2}{r^2}\left(\rho\omega^2-F^2\right)\phi=0. \label{a28}
\end{eqnarray}
This equation has been obtained previously by \cite{Wesson1978} in the  study of stability of high-temperature plasma.  Changing back to the physical variable $\xi_r$, it is easy to show that Eq. (\ref{a28}) is equivalent to
\begin{eqnarray}
\frac{1}{r}\frac{d}{dr}\left[\left(\rho\omega^2-F^2\right)r^3\frac{d \xi_r}{dr}\right]+ \left(1-m^2\right)
\left(\rho\omega^2-F^2\right)\xi_r=0. \label{CC1}
\end{eqnarray}

For the specific kink mode value of $m=1$, from Eq. (\ref{CC1}) we obtain:
\begin{eqnarray}
\frac{1}{r}\frac{d}{dr}\left[\left(\rho\omega^2-F^2\right)r^3\frac{d \xi_r}{dr}\right]=0. \label{CC2}
\end{eqnarray}
Assuming zero background magnetic twist outside the tube but an arbitrary twist inside:
\begin{equation}
\mathbf{B} = \left\{
  \begin{array}{lr}
    \left(0, B_{\varphi}\left(r\right), B_{zi}\right), & r\le a\\
    \left(0, 0, B_{ze}\right) &  r > a.
  \end{array}
\right. \label{CC3}
\end{equation}
The physical solution of Eq. (\ref{CC2}), for a trapped mode with background magnetic field Eq. (\ref{CC3}) bounded at $r=0$ and tending to $\xi_{r}=0$ as $r \rightarrow \infty$ is given as
\begin{equation}
\xi_r(r) = \left\{
  \begin{array}{lr}
    \xi_a=const, & r\le a\\
    \xi_a\left(\frac{a}{r}\right)^{2}, &  r > a.
  \end{array}
\right. \label{CC4}
\end{equation}
Physically, the radial displacement inside the tube is restricted to be constant with the radius for the $m=1$ mode (see Eq. (\ref{CC4})) or else Eq. (\ref{CC2})  would give a singularity at $r=0$. Mathematically, this singularity can be eliminated only in the case where $B_\varphi$ is not finite at the tube axis which is unphysical. The fact that Eq. (\ref{CC4}) is the same as for an untwisted tube is a new and interesting analytical result since previously \citet{Ruderman2007}, for example, only demonstrated this for the particular internal background magnetic twist of $B_{\varphi}\propto r$. Here we have shown that the radial profile of $\xi_r$ the kink mode, in the thin tube approximation, is completely independent of any prescribed background internal twist. This is also in agreement with the purely numerical study of \citet{Terradas2012} who solved the ideal linearised MHD equations in the zero-$\beta$ regime using the PDE2D code \citep{Sewell2005}. Terradas \& Goossens found that the kink mode frequency in the long wavelength approximation was not affected by their particular choice of a quadratic radial profile of $B_{\varphi}$ shown in Eq. (8) of their paper, which is consistent with our more general analytical result.

By substituting Eq. (\ref{CC4}) into the boundary condition Eq. (\ref{a27}) we obtain:
\begin{equation}
\left.\left(\rho_i+\rho_e\right)\omega^2=\left(F_{i}\left(r\right)^2+F_{e}^2\right)\right|_{r=a}-\left.2\frac{B_{\varphi}(r)F_{i}\left(r\right)}{r}\right|_{r=a}+\left.\frac{B_{\varphi}^2\left(r\right)}{r}\right|_{r=a}. \label{CC5}
\end{equation}
Here, subscripts $i$ and $e$ correspond to the internal and external parts of the magnetic flux tube correspondingly. 
By taking into account that
\begin{eqnarray}
&& F_{i}\left(r\right)=B_{\varphi}\left(r\right)/r+k_{z}B_{zi}, \nonumber \\
&& F_{e}=k_{z}B_{ze}, \label{CC6}
\end{eqnarray} 
from Eq. (\ref{CC5}) we obtain the dispersion relation
\begin{equation}
\omega^2=\frac{k_{z}^2}{\left(\rho_i+\rho_e\right)}\left(B_{zi}^2+B_{ze}^2\right),  \label{CC7}
\end{equation}
which again, is the same as the kink mode in the thin tube approximation without twist. Therefore, by Eqs. (\ref{CC7}) and (\ref{CC4}), both the frequency and $\xi_r$ eigenfunctions are unaffected by the choice of internal twist for the kink mode. 

By inspection of Eq. (\ref{CC1}) it can be seen that the case is entirely different for modes with $m \ge 2$. In such cases the  dispersion relation and $\xi_r$ eigenfunction will depend substantially on the radial profile of magnetic twist. However, unlike the $m=1$ kink mode, Eq. (\ref{CC1}) is less tractable for analysis when the radial profile of $B_{\varphi}$ is arbitrary for the modes with $m\ge 2$. Hence, in the next section we shall choose a specific radial twist profile which will enable us to do this.

\section{Internal background magnetic twist with $B_{\varphi} \propto r$}

To make analysis of Eq. (\ref{CC1}) more straightforward for fluting modes with $m \ge 2$ we will assume that inside the tube the magnetic twist varies linearly and outside it is zero, for example  \citet[][]{Bennett1999, Parker2007, Solovev2011, Solovev2012}:
\begin{equation}
\mathbf{B} = \left\{
  \begin{array}{lr}
    \left(0, B_{\varphi}\left(a\right)r/a, B_{zi}\right), & r\le a\\
    \left(0, 0, B_{ze}\right) &  r > a
  \end{array}
\right. \label{a29}
\end{equation}
where $B_{zi}$ and $B_{ze}$ are constant internal and external magnetic fields. We note that after introduction of this specific magnetic geometry (i.e. Eq. (\ref{a29})), the quantity $F\left(r\right)$ loses its dependence on $r$. 
From Eqs. (\ref{a28}) and (\ref{a29}) we obtained that inside and outside the flux,  tube Eq. (\ref{CC1}) can be represented as
\begin{equation}
(\rho \omega^2-F^2)\left[\frac{1}{r}\left(r^3\frac{d \xi_r}{d r}\right)+(1-m^2) \xi_r \right]=0.
\label{a30}
\end{equation}
By applying the boundary conditions given by Eq. (\ref{a25}) the solution of Eq (\ref{a30}), which describes perturbation of the plasma
cylinder border for $m \ge 1$, is
\begin{equation}
\xi_r = \left\{
  \begin{array}{lr}
    \xi_a\left(\frac{r}{a}\right)^{m-1}, & r\le a\\
    \xi_a\left(\frac{a}{r}\right)^{m+1}, &  r > a
  \end{array}
\right. \label{a31}
\end{equation}
Here $\xi_a=\xi_r(r=a)$.
By substituting Eq. (\ref{a31}) into Eq. (\ref{a27}) we obtain the dispersion relation in the form:
\begin{eqnarray}
&& \omega^2=\frac{1}{\left(\rho_i+\rho_e\right)}\left[k_{z}^2\left(B_{zi}^2+B_{ze}^2\right)+m\left(m-1\right)\frac{B_{\varphi}^2(a)}{a^2}\right.+ \nonumber \\
&& \left. \frac{2k_z}{a}\left(m-1\right)B_{\varphi}(a)B_{zi}\right]. \label{a32}
\end{eqnarray}
This dispersion relation is valid for positive azimuthal $m$ and longitudinal $k_z$ wave numbers. It is important to note that this relation is invariant under the substitution $(m, k_z) \rightarrow (-m, -k_z)$, resulting in:

\begin{eqnarray}
&& \omega^2=\frac{1}{\left(\rho_i+\rho_e\right)}\left[k_{z}^2\left(B_{zi}^2+B_{ze}^2\right)+m\left(m-sign(m)\right)\frac{B_{\varphi}^2(a)}{a^2}\right.+ \nonumber \\
&& \left. \frac{2k_z}{a}\left(m-sign(m)\right)B_{\varphi}(a)B_{zi}\right]. \label{a32.1}
\end{eqnarray}
Hence, Eq. (\ref{a32.1}) describes wave propagation in both directions along the twisted magnetic tube. If current is absent, meaning that $B_{\varphi i} =0$, the two last terms vanish and Eq. (\ref{a32.1}) describes the kink mode  \citep[see e.g.][]{Ryutov1976, Spruit1982, Edwin1983}. Although $m$ in Eq (\ref{a32.1}) can either be positive or negative, in the following sections we analyse only the case when $m>0$.

   \begin{figure}
   \centering
{\tiny }   \includegraphics[angle=-90, origin=c, width=8cm]{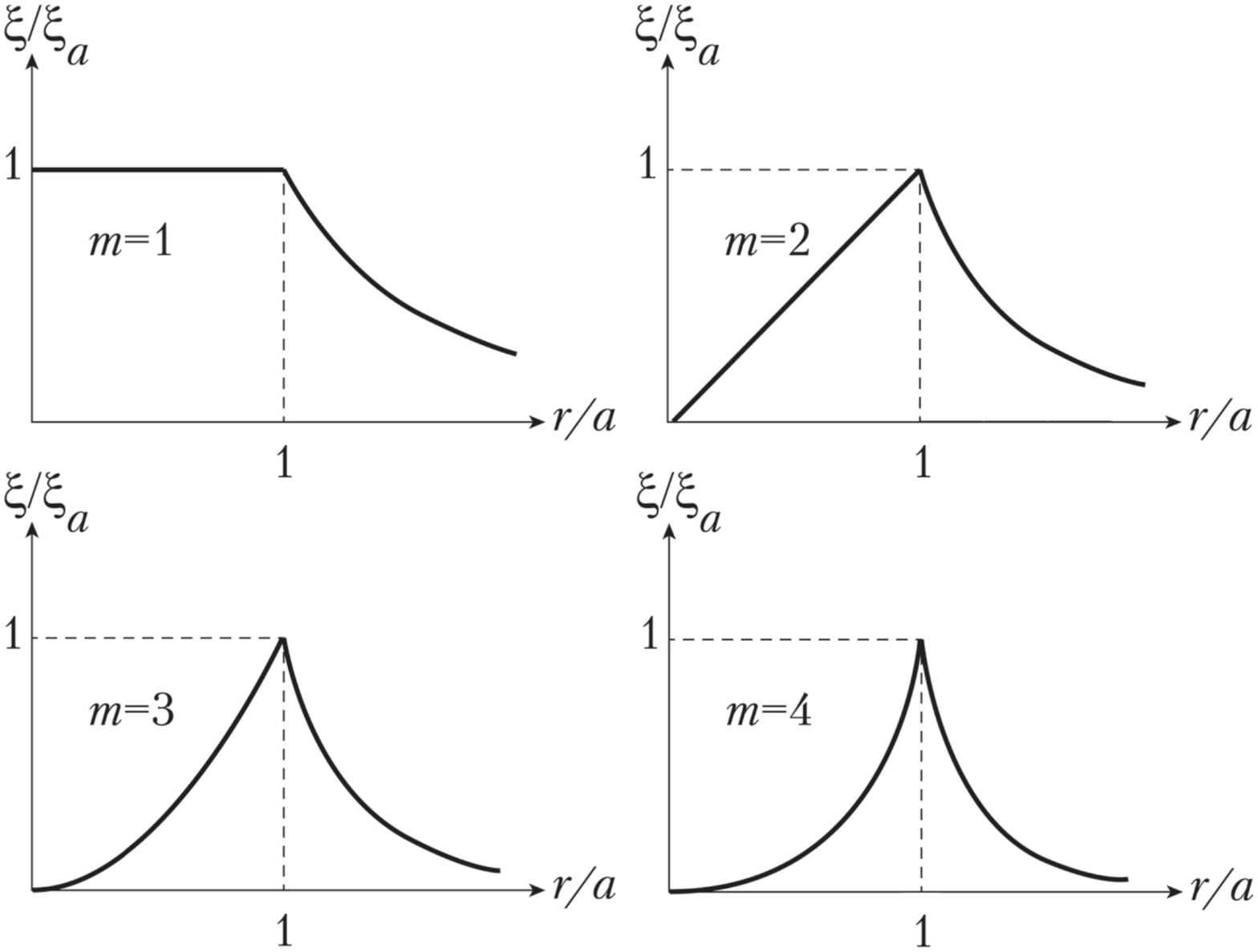}
      \caption{Radially dependent eigenfunctions $\xi_r(r)$ for axially symmetric magnetic flux tube with a free boundary 
      are shown for azimuthal wave numbers $m=1$, 2, 3, and 4.
              }
         \label{Fig2}
   \end{figure}
\begin{equation}
\xi_r(r) = \left\{
  \begin{array}{lr}
    \xi_a=const, & r\le a\\
    \xi_a\left(\frac{a}{r}\right)^{2}, &  r > a,
  \end{array}
\right. \label{a34}
\end{equation}
where $\xi_a=\xi_r\left(r=a\right)=\phi_a/r$. This function is constant up to the boundary of the flux tube and then decreasing
to the infinitely small values as $r \rightarrow \infty$ (see Fig.~\ref{Fig2}).

The square of the frequency of fluting modes ($m \ge 2$) given by Eq (\ref{a32}), in contrast to kink modes ($m=1$), has a minimum value when
\begin{equation}
k_z=-\frac{\left(m-1\right)}{a}\frac{B_{\varphi}(a)B_{zi}}{B_{zi}^2+B_{ze}^2} \nonumber
\end{equation}
and is equal to
\begin{equation}
\omega_{min}^2=\frac{B_{\varphi}^2(a)}{a^2}\frac{\left(m-1\right)}{\left(\rho_i+\rho_e\right)}\left[1+\frac{\left(m-1\right)B_{ze}^2}{B_{zi}^2+B_{ze}^2}\right]. \label{a35}
\end{equation}
In contrast to the kink mode, Eq. (\ref{a35}) shows that a low frequency cut-off is introduced for fluting modes due to background twist. For all $m \ge 2$ modes the eigenfunction $\xi_r$ has a form of power function Eq. (\ref{a31}) and describes perturbations which are localised at the surface of the twisted magnetic flux tube (see Fig.~\ref{Fig2}). It is also interesting to note that the sign of the $\xi_r$ eigenfunctions $\xi_r$ do not vary as $r$ increases.

\section{Comparison with previous results}

Equation (\ref{a29}) describes the same background magnetic field configuration as \citet{Bennett1999}, to allow for direct comparison. In the paper by Bennett et al. the authors obtain the specific dispersion relation which describes sausage ($m=0$) waves in an incompressible magnetic flux tube with uniform twist embedded within an untwisted magnetic environment, that is:
\begin{eqnarray}
&& \frac{\left. \left(\omega^2-\omega_{A}^2\right)\frac{x}{I_{m}(x)}\frac{dI_{m}(x)}{dx}\right|_{x=\left|m_0\right|a}-\frac{2mB_{\varphi}(a)\omega_{Ai}}{a\sqrt{\rho_i}}}{{\left(\omega^2-\omega_{Ai}^2\right)^2-4\omega_{Ai}^2}\frac{B_{\varphi}^2(a)}{a^2\rho_{i}}}= \nonumber \\
&& \frac{\left.\frac{y}{K_{m}(y)}\frac{dK_{m}(y)}{dy}\right|_{y=\left|k_z\right|a}}{\frac{\rho_e}{\rho_i}\left(\omega^2-\omega_{Ae}^2\right)+\left. \frac{B_{\varphi}^2(a)}{a^2\rho_i}\frac{y}{K_{m}(y)}\frac{dK_{m}(y)}{dy}\right|_{y=\left|k_z\right|a}}. \label{a38}
\end{eqnarray}
Here $I_m(x)$ and $K_m(x)$ are modified Bessel functions of the first and second kind with an imaginary argument,
\begin{equation}
 m_{0}^2a^2=k_{z}^2a^2\left(1-\frac{4B_{\varphi}^2}{r^2}\frac{F^2}{\left(\rho\omega^2-F^2\right)^2}\right) \label{a39} 
 ,\end{equation}
 and
 \begin{equation}
 \omega_{A\alpha}^2=\frac{1}{\rho_\alpha}\left(\frac{m}{a}B_{\varphi}(a)+k_zB_{z\alpha}\right)^2.
\end{equation}  
The subscript $\alpha$ refers to $i$ and $e$, which corresponds to the quantities either inside and outside of the tube, respectively.
We note that \citet{Bennett1999} have shown numerically (see. Fig. 5 of their paper) that for 
the kink mode ($m=1$), the phase and group velocities are approximately equal, such that:
\begin{equation}
\frac{\omega}{k_{z}}\approx \frac{\partial \omega}{\partial k_{z}}\approx \left(\frac{B_{zi}^2+B_{ze}^2}{\rho_i+\rho_e}\right)^{\frac{1}{2}}, \label{aa1}
\end{equation}
which is in a great agreement with Eq. (\ref{CC7}). 
We now want to check that the dispersion relations given by Eq. (\ref{a32}) and Eq. (\ref{a38}) are equivalent to each other in the long wavelength approximation when $m=0$ and furthermore, that the eigenfunctions are also the same. As mentioned in \citet{Bennett1999}, it is difficult to analyse analytically the modes with $m\ge1$ and hence this was not attempted in their work. This is partly due to the fact that modes with $m\ge1$ can exist with both $m_{0}^2>0$ and $m_{0}^2<0$. Equation (\ref{a38}) was obtained for the particular case of $m_{0}^2>0$. Let us show that these conditions do not actually affect the final form of the dispersion equation in the long wavelength approximation.  In the case in which  $m_{0}^2<0$ only one change in the left hand side of Eq. (\ref{a38}) will appear due to the replacement (see.
Appendix B, Eq. (\ref{AA8}))
\begin{equation}
\left. \frac{x}{I_m}\frac{dI_m}{dx}\right|_{x=m_{0}a} \rightarrow \left. \frac{x}{J_m}\frac{dJ_m}{dx}\right|_{x=\left| m_{0}\right|a} \label{aa2}
\end{equation}
Here $J_{m}(x)$ is the Bessel function of the first kind. In the long-wave approximation  $\left|m_0\right| a \approx \left|k_z\right| a \ll 1$, therefore the Bessel functions satisfy following relations:
\begin{equation}
\left. \frac{xI_{m}^{\prime}}{I_{m}}\right|_{x=m_0 a \ll 1} = \left. \frac{xJ_{m}^{\prime}}{J_{m}}\right|_{x=\left|m_0\right| a \ll 1} =
-\left. \frac{xK_{m}^{\prime}}{K_{m}}\right|_{x=\left|k_z\right| a \ll 1} =m. \label{a40}
\end{equation}
By applying relations shown in Eq. (\ref{a40}) we arrive to the conclusion that, independent of the sign of $m_{0}^2$, Eq. (\ref{a38}) can be written in the form:
\begin{equation}
\left(\omega^2-\omega_{Ai}^2\right)+2\omega_{Ai}\left(\frac{B_{\varphi}\left(a\right)}{a\sqrt{\rho_i}}\right)+\frac{\rho_e}{\rho_i}\left(\omega^2-\omega_{Ae}^2\right)-m\frac{B_{\varphi}^2(a)}{a^2\rho_i}=0. \label{a41}
\end{equation}
After some algebra it can be shown that Eq. (\ref{a41}) coincides with the previously obtained dispersion relation in Eq. (\ref{a32}).

We now wish to further confirm our results by comparing the eigenfunctions of Eqs. (\ref{a31}) and (\ref{a34}) with those of \citet{Bennett1999}. Bennett et al. obtained a solution (see (\ref{AA6})) for the total pressure perturbation $\delta p_1$  when $m_{0}^2>0$:
\begin{equation}
\delta p_1 = \left\{
  \begin{array}{lr}
    I_m\left(m_0 r\right), & r\le a\\
    K_m\left(\left|k_z\right|r\right), &  r > a.
  \end{array}
\right.
 \label{aa3}
\end{equation}
If $m_{0}^2<0$, the expression for $\delta p_1$ has the following form:
\begin{equation}
\delta p_1 = \left\{
  \begin{array}{lr}
    J_m\left(\left|m_0\right|r\right), & r\le a\\
    K_m\left(\left|k_z\right|r\right), &  r > a.
  \end{array}
  \right.
 \label{aa4}
\end{equation}
Since in the long wavelength approximation, functions $I_m (x)$, $J_m (x)$ and $K_m (x)$ satisfy relations:
\begin{equation}
I_m (x), \, \left. J_m (x)\right|_{x \ll 1} \sim x^m, \, \left. K_{m}(x)\right|_{x \ll 1}\sim x^{-m},
\end{equation}
$\delta p_1$, independent of the sign of $m_{0}^2$, is given by:
\begin{equation}
\delta p_1 \sim \left\{
  \begin{array}{lr}
    r^m, & r\le a\\
    r^{-m}, &  r > a.
  \end{array}
  \right.
 \label{a42}
\end{equation}
According to (\ref{AA1}), (\ref{AA4}) and (\ref{AA5}) the total pressure perturbation $\delta p_1$ is associated with radial
displacement $\xi_r$ as:
\begin{equation}
r\xi_r=const_1r\frac{d}{dr}\delta p_1 +const_2 \delta p_1. \label{a43}
\end{equation}
From Eqs. (\ref{a42}) and (\ref{a43}) the radial displacement $\xi_r$ depends on $r$ as follows:
\begin{equation}
\xi_r \sim \left\{
  \begin{array}{lr}
    r^{m-1}, & r\le a\\
    r^{-m-1}, &  r > a.
  \end{array}
  \right.
\label{aa5}
\end{equation}
This is consistent with the radial dependence of the eigenfunctions  previously derived from dispersion relation (\ref{a32}) and shown in Eqs. (\ref{a31}) and (\ref{a34}).
 
Now we shall compare our results with those obtained by \citet{Ruderman2007} who studied linear non-axisymmetric oscillations of a thin magnetic tube in presence of the weak internal magnetic twist, that is $B_{\varphi}(r\le a)\ll B_{zi}$ and $B_{\varphi}(r>a)=0$. Ruderman obtained the dispersion relation for incompressible MHD perturbations in the form of 
\begin{equation}
\omega^2=\frac{2B_{0}^2}{\left(\rho_i+\rho_e\right)}\left\{ k_{z}^2+\frac{A\left(m-sign\left(m\right)\right)}{2B_{0}^2}\left(Am+2B_{0}k_z\right)\right\}.  \label{aa6}
\end{equation}
It can be seen that this equation is the same as Eq. (\ref{a32.1}) when $B_{\varphi}(a)/a=A$ and $B_{zi}=B_{ze}=B_{0}$. 

Therefore, the dispersion relation of Eq. (\ref{a32.1}) and the resulting expressions for the eigenfunctions in Eqs. (\ref{a31}) and (\ref{a34}) are in excellent agreement with the previous results of both  \citet{Bennett1999} and \citet{Ruderman2007}, who studied more specialised cases of our more general plasma and magnetic field background configuration.

\section{Conclusions}

In this work we analysed the incompressible linear MHD modes of a twisted magnetic flux tube. Special attention was given to the problem of finding the eigenvalues and eigenfunctions of modes with $m\ge 1$ in the long wavelength limit. The description of these modes leads to significant mathematical difficulties as noted in \citet{Bennett1999}. However, in this current work, by using the long wavelength approximation we have made significant analytical progress. Notably, we have shown that the dispersion relation given in Eq. (\ref{CC7}) for the $m=1$ kink mode is completely unaffected by the radial profile of background internal magnetic twist. It was also found from the $\xi_r$ eigenfunction that the kink mode is body-like in character (see Eq. (\ref{a34})) and the higher order modes ($m \ge 2$) are surface-like (see Eq. (\ref{a31})). Hence, these results are the same as known previously in the case of no background magnetic twist for the particular choice of having $B_{\varphi} \propto r$ inside the tube. However, it was shown in Eq. (\ref{a35}), that a low frequency cut-off was introduced for fluting modes due to the presence of background twist, in contrast to the cut-off free propagation for the kink mode.

It can be seen by the derived dispersion relation shown in Eq. (\ref{CC7}) that the phase speed of the kink mode, equal to the group speed in the long wavelength limit, will not be affected at all by the presence of internal background magnetic twist. When there is no internal twist, all surface modes tend to the kink speed in the long wavelength limit. However, this is not the case when twist is present, since it can either work for or against the propagation speed of these modes. For example, when $m$, $k_z$ and $B_{\varphi}$ are all the same sign, the resulting phase speed is increased relative to the kink speed. If the sign of $B_{\varphi}$ is opposite to that of $m$ and $k_z$, then the speed is reduced relative to the kink speed. This is analogous to MHD wave mode propagation along a magnetic flux tube in the presence of field-aligned flow. That is to say, if the wave is travelling in the same or opposite direction to that of the flow, the speed is increased/decreased relative to the case when no flow is present. Hence, this presents a challenge in interpreting what differences in observed counter-propagating MHD wave-mode speeds could be caused by, meaning a magnetic twist, field-aligned flow, or a combination of both. Futhermore, to our knowledge $m \ge 2$ modes have still yet to be identified in solar atmospheric observations of thin twisted magnetic flux tubes, for example, chromospheric fibrils, mottles, and spicules. However, the current work suggests their sensitivity to magnetic twist would make them a very interesting future case study.

\begin{acknowledgements}
All authors thank the referee and Prof. M. Ruderman for the 
constructive suggestions that improved the paper. 
OC and AK would like to thank the Ukrainian Scientific and Technical Center, PN 6060; Integrated Scientific 
Programmes of the National Academy of Science of Ukraine on Space Research and Plasma Physics. 
VF, GV thank the STFC and  Royal Society-Newton Mobility Grant
for the support received. VF also thanks to Newton Fund MAS/CONACyT Mobility
Grants Program.
\end{acknowledgements}

\begin{appendix}
\section{Obtaining an expression for $\delta p_{1}$}

From the following equations:
 \begin{eqnarray}
 && \delta p_1=\delta p + \mathbf{B}\cdot \delta \mathbf{B}, \nonumber \\
 && \delta p=-\bfxi \cdot \mathbf{\nabla} p -\gamma p \bfbigtriangledown \cdot \bfxi, \label{BB1} \\
 && \delta \mathbf{B}=\bfbigtriangledown \times \left[\bfxi \times \mathbf{B}\right]=\left(\mathbf{B}\cdot\bfbigtriangledown \right)\bfxi-\left(\bfxi\cdot\bfbigtriangledown\right)\mathbf{B}-\mathbf{B}\left(\mathbf{B}\cdot\bfxi\right). \nonumber 
\end{eqnarray}
By using the equation of equilibrium
\begin{equation}
\bfbigtriangledown p=\left(\mathbf{B}\cdot \bfbigtriangledown\right)\mathbf{B}-\bfbigtriangledown\left(B^2/2\right), \label{BB11} \\
\end{equation}
after some algebra from Eq. (\ref{BB1}) we obtain
\begin{equation}
\delta p_1=-\bfxi\cdot\left(\mathbf{B}\cdot\bfbigtriangledown\right)\mathbf{B}-\gamma p \left(\bfbigtriangledown\cdot\bfxi\right)+\mathbf{B}\cdot\left(\mathbf{B} \cdot \bfbigtriangledown \right)\bfxi-
B^2\bfbigtriangledown\cdot\bfxi. \label{BB2} 
\end{equation}
By using the vector relations 
\begin{eqnarray}
&& -\bfxi\cdot\left(\mathbf{B}\cdot \bfbigtriangledown\right)\mathbf{B}+\mathbf{B}\cdot\left(\mathbf{B}\cdot \bfbigtriangledown\right)\bfxi=-2\bfxi\cdot\left(\mathbf{B\cdot\bfbigtriangledown}\right)\mathbf{B}+ \nonumber \\
&& \left(\mathbf{B}\cdot\bfbigtriangledown\right)\left(\bfxi\cdot\mathbf{B}\right)=-2\bfxi\cdot\left(\mathbf{B}\cdot\bfbigtriangledown\right)\mathbf{B}+\bfbigtriangledown\cdot\left[\mathbf{B}\left(\bfxi\cdot\mathbf{B}\right)\right]= \nonumber \\
&& -2 B\bfxi\cdot\left(\mathbf{\tau}\cdot\bfbigtriangledown\right)\mathbf{\tau}B+\bfbigtriangledown\cdot \left[B^2\left(\bfxi\cdot\mathbf{\tau}\right)\tau\right], \label{BBB12}
\end{eqnarray}
where $\tau=\mathbf{B}/B$ we can rewrite Eq. (\ref{BB2}) as
\begin{eqnarray}
&& \delta p_1= -2B\bfxi\cdot\left(\tau\cdot\bfbigtriangledown\right)\tau B+\bfbigtriangledown\cdot\left[B^2\left(\bfxi\cdot\tau\right)\right]- \nonumber \\
&& \gamma p\bfbigtriangledown\cdot \bfxi - B^2\bfbigtriangledown\cdot \bfxi. \label{BB3}
\end{eqnarray}
The first term on the RHS of Eq. (\ref{BB3}) can be transformed as follows:
\begin{eqnarray}
&&-2B\bfxi\cdot\left(\tau\cdot\bfbigtriangledown\right)\tau B =-2B^2\bfxi\cdot\left(\tau\cdot \bfbigtriangledown\right)\tau- \nonumber \\
&& 2\left(\bfxi\cdot\tau\right)B\left(\tau\cdot\bfbigtriangledown B\right)=-2B^2\bfxi\cdot\bfkappa-2\xi_{\parallel}B\left(\tau\cdot\bfbigtriangledown B\right)= \nonumber \\
&&-2B^2\bfxi\cdot\bfkappa-\xi_{\parallel}\left(\tau\cdot\bfbigtriangledown B^2\right). \label{BB4}
\end{eqnarray}
Here $\bfkappa=\left(\tau\cdot\bfbigtriangledown\right)\tau$ is the vector of curvature of the magnetic field lines. To obtain Eq. (\ref{BB4}) we take into account that displacement vector $\bfxi$ is represented in the form $\bfxi=\bfxi_{\perp}+\xi_{\parallel}\mathbf{\tau}$, where subscripts $\perp$ and $\parallel$ correspond to the perpendicular and parallel components to the equilibrium magnetic field. Therefore, Eq (\ref{BB3}) can be rewritten in the form: 
\begin{eqnarray}
&& \delta p_1 =-2B^2\bfxi\cdot\bfkappa-\xi_{\parallel}\left(\tau\cdot\bfbigtriangledown B^2\right)+\bfbigtriangledown \cdot\left(\xi_{\parallel} B^2 \tau\right)- \nonumber \\
&& \gamma p \bfbigtriangledown\cdot\xi -B^2 \bfbigtriangledown \bfxi. \label{BB5}
\end{eqnarray}
From the second and third terms in Eq. (\ref{BB5}) we can obtain
\begin{eqnarray}
&&-\xi_{\parallel}\left(\tau\cdot\bfbigtriangledown B^2\right)+\bfbigtriangledown\cdot\left(\xi_{\parallel} B^2 \mathbf{\tau}\right)=-\xi_{\parallel}\left(\tau\cdot \bfbigtriangledown B^2\right)+ \nonumber \\
&&\xi_{\parallel}\left(\tau\cdot B^2\right)+ B^2 \bfbigtriangledown\left(\xi_{\parallel} \mathbf{\tau}\right)=B^2\bfbigtriangledown \left(\xi_{\parallel}\mathbf{\tau}\right). \label{BB6}
\end{eqnarray}
From Eqs. (\ref{BB5}) and (\ref{BB6}) it follows that:
\begin{eqnarray}
&& \delta p_1=-2B^2\bfxi\cdot\bfkappa+B^2\bfbigtriangledown\cdot\left(\xi_{\parallel}\tau\right)-\gamma p \bfbigtriangledown\cdot \bfxi - \nonumber \\
&& B^2\bfbigtriangledown\bfxi =-2B^2\bfxi\cdot\bfkappa-\gamma p\bfbigtriangledown\cdot \bfxi -B^2\bfbigtriangledown\cdot \bfxi_{\perp}. \label{BB7}
\end{eqnarray}
The first term on the RHS of Eq (\ref{BB7}) is  equal to:
\begin{eqnarray}
&& -2B^2\bfxi\cdot\bfkappa=-2B^2\left(\bfxi_{\perp}+\bfxi_{\parallel}\tau\right)\left(\tau\cdot\bfbigtriangledown\right)\mathbf{\tau}= \nonumber \\
&& -2B^2\bfxi_{\perp}\cdot\bfkappa-2B^2\bfxi_{\parallel}\mathbf{\tau}\cdot\left(\mathbf{\tau}\cdot \bfbigtriangledown\right)\mathbf{\tau}= \nonumber \\
&& -2B^2\bfxi_{\perp}\cdot\bfkappa-B^2\bfxi_{\parallel}\left(\mathbf{\tau}\cdot\bfbigtriangledown\right)1=-2B^2\bfxi_{\perp}\cdot\bfkappa.
\end{eqnarray}
Finally, we obtain the second expression in Eq. (\ref{a2}), that is, 
\begin{equation}
\delta p_1=-\gamma p \bfbigtriangledown\cdot \bfxi-B^2\left(\bfbigtriangledown\cdot\bfxi_{\perp}+2\bfxi_{\perp}\cdot\bfkappa\right). \label{BB8}
\end{equation}

\section{Exact dispersion relation for the magnetic field described as (\ref{a29}) for $m_{0}^2<0$}
To obtain the exact dispersion relation for the case (\ref{a29}) we have used the following set of Eqs: (\ref{a14}) and (\ref{a20}), (\ref{a21}) and (\ref{a22}). From Eqs. (\ref{a20}) and (\ref{a22}) we have:
\begin{eqnarray}
&& \rho\left(\omega^2-\omega_{A}^2\right)\frac{d\delta p_1}{dr}=2\frac{m}{r^2}B_{\varphi}\left(\frac{m}{r}B_{\varphi}+k_{z}B_{z}\right)\delta p_1+ \nonumber \\
&& \xi_{r}\left\{\left[\rho\left(\omega^2-\omega_{A}^2\right)\right]^2+2\rho B_{\varphi}\left(\omega^2-\omega_{A}^2\right)\frac{d}{dr}\left(\frac{B_{\varphi}}{r}\right)-\right. \nonumber \\
&& \left. 4\rho \omega_{A}^2\left(\frac{B_{\varphi}^2}{r^2}\right)\right\}. \label{AA1}
\end{eqnarray}
Equation (\ref{AA1}) together with Eqs. (\ref{a20}) and (\ref{a14}) represent a system of equations
with respect to $\xi_r$ and $\delta p_1$. From this system it is easy to obtain the equation for $\delta p_1$:
\begin{eqnarray}
&& \frac{d^2}{dr^2}\delta p_1+\left[\frac{C_3}{rD}\frac{d}{dr}\left(\frac{rD}{C_3}\right)\right]\frac{d}{dr}\delta p_1+
\left[\frac{C_3}{rD}\frac{d}{dr}\left(\frac{rC_1}{C_3}\right)\right.+ \nonumber \\
&& \left. \frac{1}{D^2}\left(C_2C_3-C_{1}^2\right)\right]\delta p_1=0, \label{AA2}
\end{eqnarray}
where
\begin{eqnarray}
&& D=\rho\left(\omega^2-\omega_{A}^2\right), \, C_1=-2\frac{mB_{\varphi}}{r^2}\left(\frac{m}{r}B_{\varphi}+k_{z}B_{z}\right), \nonumber \\
&& C_2=-\left(\frac{m^2}{r^2}+k_{z}^2\right),  \nonumber \\
&& C_3=D^2+2DB_{\varphi}\frac{d}{dr}\left(\frac{B_{\varphi}}{r}\right)-4\rho\omega_{A}^2\left(\frac{B_{\varphi}^2}{r^2}\right). \label{AA3}
\end{eqnarray}
From (\ref{AA3}) it follows that in for the magnetic field (\ref{a29}) are following relations are satisfied:
\begin{equation}
\rho\omega_{A}^2=const, \, \frac{B_{\varphi}}{r}=const, \, \frac{m}{r}B_{\varphi}+k_z B_z=const. \label{AA4}
\end{equation}
Therefore,
\begin{equation}
D=const, \, rC_1=const, \, C_3=const. \label{AA5}
\end{equation}
By taking into account (\ref{AA4}) and (\ref{AA5}), from (\ref{AA2}) we obtain
\begin{eqnarray}
&& \frac{d^2}{dr^2}\delta p_1+\frac{1}{r}\frac{d}{dr}\delta p_1-\left(\frac{m^2}{r^2}+k_{z}^2\right)\delta p_1+ \nonumber \\
&& \frac{4\left(B_{\varphi}^2/r^2\right)\omega_{A}^2k_{z}^2}{\rho\left(\omega^2-\omega_{A}^2\right)^2}\delta p_1=0. \label{AA6}
\end{eqnarray}
Equation (\ref{AA6}) coincides with the Eq. (13) of work \citet{Bennett1999} and satisfies the boundary
conditions (\ref{a25}) and (\ref{a27}). 

Following \citet{Bennett1999}, we introduce $m_{0}^2$ (see (\ref{a39})). We assume $m_{0}^2$<0, that gives a solution (\ref{AA6}) in the form
\begin{equation}
\delta p_1 = \left\{
  \begin{array}{lr}
    A_i J_m\left(\left|m_0\right|r\right), & r\le a\\
    A_e K_m\left(\left|k_z\right|r\right), &  r > a
  \end{array}
\right.  
 \label{AA7}
\end{equation}
To obtain (\ref{AA7}) we take into account that $m_0=\left|k_z\right|$ for $r>a$. From (\ref{AA1}), (\ref{a25}), (\ref{a27}) and (\ref{AA7}), we obtain desired dispersion relation:
\begin{eqnarray}
&& \frac{\left. \left(\omega^2-\omega_{A}^2\right)\frac{x}{J_{m}(x)}\frac{dJ_{m}(x)}{dx}\right|_{x=\left|m_0\right|a}-\frac{2mB_{\varphi}(a)\omega_{Ai}}{a\sqrt{\rho_i}}}{{\left(\omega^2-\omega_{Ai}^2\right)^2-4\omega_{Ai}^2}\frac{B_{\varphi}^2(a)}{a^2\rho_{i}}}= \nonumber \\
&& \frac{\left.\frac{y}{K_{m}(y)}\frac{dK_{m}(y)}{dy}\right|_{y=\left|k_z\right|a}}{\frac{\rho_e}{\rho_i}\left(\omega^2-\omega_{Ae}^2\right)+\left. \frac{B_{\varphi}^2(a)}{a^2\rho_i}\frac{y}{K_{m}(y)}\frac{dK_{m}(y)}{dy}\right|_{y=\left|k_z\right|a}} \label{AA8}
\end{eqnarray}
Used in (\ref{AA8}) variables and parameters are explained in Eq. (\ref{a39}).

\end{appendix}

\end{document}